\documentstyle[11pt, aaspp4]{article}

\begin{document}

\title{Mid-Infrared Imaging of Alpha Orionis}
\author{S. A. Rinehart, T. L. Hayward, J. R. Houck}
\affil{Cornell University}
\authoraddr{208 Space Sciences, Cornell University, Ithaca NY 14853}
\authoremail{rinehart@tristan.tn.cornell.edu}

\begin{abstract}

We have imaged the M2Iab supergiant star $\alpha$~Ori at 11.7 \micron
\ and 17.9 \micron \ using the Hale 5-m telescope and an array
camera with sub-arcsecond angular resolution.  Our images reveal the
circumstellar dust shell extending out to $\sim$ 5$\arcsec$ from the
star.  By fitting two multi-parameter models to the data, we find
values for the dust shell inner radius, the temperature at this
radius, the percentage flux produced by the circumstellar material at
each of our wavelengths, and the radial dependence of the dust mass
density.  We present profiles of the mean surface brightness as a
function of radius and the power detected from within annular rings as
a function of radius.

\end{abstract}


\section{Introduction}

As stars ascend the asymptotic giant branch (AGB), they shed mass,
developing a circumstellar shell consisting of atomic gas, molecular
gas, and dust.  One of the most frequently studied AGB stars
is the oxygen-rich M2 Iab star $\alpha$ Ori, which is relatively
easily observed due to its high luminousity and proximity, $d \simeq
130$~pc (where 1$\arcsec =$ 130 AU).  Observations of its spatially
extended circumstellar shell have been made in many wavelength bands.
Mapping at 21 cm has shown that H I emission extends $\sim 1\arcmin$
from the star (\cite{BK}).  Large beam far-IR spectra have detected [C I] 609
$\mu$m in the shell (\cite{HBCF}) as well as [O I] 63 $\mu$m and [Si
II] 35 $\mu$m from what is believed to be the region between the
chromosphere and the extended shell (\cite{HG}).  In the 10 $\mu$m
band, the shell has been observed using both interferometric
(\cite{BDTD}) and drift scan (\cite{BTV}) techniques, and long-slit
7--14 $\mu$m spectra have revealed spatially extended emission from silicate
grains (\cite{SGL}).

Despite the quantity of previous observations, there are still
many unanswered questions about the morphology and the optical
properties of the dust.  The drift scans taken by Bloemhof et
al. (1984) indicate an asymmetry in the dust shell, implying possible
smaller scale structure in the circumstellar envelope; however, no
fully sampled, sub-arcsecond resolution map of the shell has been
published.  In addition, the size, temperature distribution, and dust
condensation radius of the silicate grains have not been determined.

This paper reports the first successful attempt to spatially resolve
the dust cloud surrounding $\alpha$ Ori by direct imaging in the
mid-IR.  We also present several physical models fit to our
observations.

\section{Observations and Data Reduction}

The observations were made using SpectroCam-10 (\cite{SCam}) at the
Cassegrain focus of the Hale 5-m telescope.\footnote{Observations
at the Palomar Observatory were made as part of a continuing
collaborative agreement between the California Institute of
Technology, the Jet Propulsion Laboratory, and Cornell University.}
The angular size of the 128x128 Si:As BIB detector's pixels
(0$\farcs$256) well samples the diffraction limit of the telescope
(0$\farcs$5 FWHM at 10 $\mu$m, 0$\farcs$9 FWHM at 18 $\mu$m),
making it possible to resolve the dust shell of the star in the
mid-IR.  Experience has shown that the diffraction limit is frequently
achieved with this system.  The observations were made on 1994
November 13 and 15.  Data were taken using 11.7 $\mu$m and 17.9 $\mu$m
filters with bandwidths of $\sim$ 1 $\mu$m and $\sim$0.5 $\mu$m,
respectively.  The 11.7 $\mu$m filter was used for observations on
both nights, while the 17.9 $\mu$m filter was used only on the second.
Observations of $\alpha$ Tau were interleaved with the observations of
$\alpha$ Ori on both nights, in order to monitor the point spread
function over the course of our observations.  An active tip-tilt
system was employed, minimizing seeing degradation due to telescope
movement, and all observations were made with standard beam-switched
chopping and nodding techniques.

Individual exposures of $\alpha$ Ori and $\alpha$ Tau were
flat-fielded, coadded, and mosaiced to produce images of $\alpha$~Tau
and $\alpha$~Ori at each observed wavelength.  In the mosaicing
process, each input image was rebinned by a factor of 3, replacing
each pixel with 9 pixels (3$\times$3).  In this magnification, no
interpolation was used; each of the subpixels was assigned a flux
equal to the one ninth of the original pixel's flux value.  The images
were registered to the nearest subpixel (0$\farcs$083) and coadded.
The final image of $\alpha$~Ori at 11.7\micron \ consists of 11
coadded frames (total on-source integration time of 26.3 s), and the
final image of $\alpha$~Ori at 17.9\micron \ consists of 9 coadded
frames (21.7 s on-source integration).  The magnitude of $\alpha$ Tau
at each wavelength was determined by interpolating between published values
listed by Gezari et al. (1993) between 2.2 and 22 $\mu$m, and
flux-magnitude relations were derived from Cohen et al. (1992).  These
were used to flux calibrate the images of $\alpha$ Ori at both
wavelengths.

\section{Analysis}

The 11.7~\micron\ and 17.9~\micron\ data are presented in
Figures 1 and 2.   Panels (a) and (b) of each figure show the 
$\alpha$~Ori and $\alpha$~Tau mosaics.  Figures 1c and 2c display
the total fluxes within a series of annulli of radius $b$ and 
width $w = 1/12 \arcsec$ around the stars, with the $\alpha$~Tau
profiles scaled to match the $\alpha$~Ori photospheric emission.
The extra emission from the $\alpha$~Ori shell is clearly seen
in both plots.

The mid-infrared emission from $\alpha$ Ori arises from two physically
distinct components: the stellar photosphere with a diameter of
0$\farcs$044 (Dyck et al. 1992), which is much smaller than our
diffraction limit, and the extended circumstellar material.  Our
images represent the superposition of these two components as
convolved with the instrumental point-spread function (PSF).
In principle we can generate an image of the shell by subtracting
the point source component, but in practice it is difficult to
determine what fraction of the total flux is due to the point source. 
As an example, Figures 1d and 2d show $\alpha$~Ori after subtracting PSF's 
derived by scaling the $\alpha$~Tau images to contain 33\% 
of the total (star + shell) $\alpha$~Ori flux at 11.7~\micron \ and
45\% at 17.9~\micron.  (These scaling factors were derived from
the spike+ modelling in \S 3.3.)  We found that for $b < 1$ arcsec, 
such subtractions were very sensitive to the scaling factor, the
stability of the seeing (which affected the width of the PSF),
and the optical aberrations in the telescope (which caused a
three-lobed effect that varied slightly between the two stars,
especially apparent at 11.7~\micron).  Because the photosphere emission
is so bright, small errors in its removal have drastic effects
on the remaining faint, extended component.  However, at $b > 1$ arcsec,
where the photospheric contribution is relatively small, 
the subtraction results were relatively insensitive
to the choice of scaling factor. 

Both the 11.7 and 17.9~\micron\ images reveal that at $b > 1$ arcsec the 
shell flux is distributed nearly symmetrically about the star.
Unfortunately, the 17.9~\micron\ images of both stars contained a filter
reflection (at 4\% of the flux of the star) to the SE of the star.  
To reduce the effect of this reflection, we did not use the SE 
quadrant in computing our 17.9~\micron\ radial profiles for the
modelling described below.

We first attempted to use maximum entropy and maximum likelihood 
deconvolution techniques in our analysis, but they did not produce
stable results for the low surface brightness shell because of the
extremely bright nearby point source. Therefore, we developed a series
of multi-parameter models for the shell emission and numerically
fitted them to the observational data.

\subsection{3--D Models}

The numerical models we developed for $\alpha$ Ori assume the basic
geometry shown in Figure 3.  The star is a a 3600 K blackbody point
source.  The shell is optically thin to the photospheric radiation with dust temperature given by $T(r) = T_o
(r_o/r)^{1/2}$, where $r$ is the distance from the star and $T_o$ is
the temperature at the reference radius $r_o = 1\arcsec$.  The mass
density of the dust is taken as $\rho = \rho (r)$; the
functional form is dependent upon the model geometry.  The dust
emissivity $\epsilon(\lambda)$ is the astronomical silicate function
from Draine \& Lee 1984.  The total flux along a line of sight at
projected radius $b$ equals the sum of the stellar and integrated dust
emission:
 
\begin{equation}
I_{\lambda}(b)_{\alpha \scriptstyle Ori \mit} = {I_{\lambda}(b)_{\scriptstyle phot \mit} + p \frac{2hc^2}{\lambda ^5}\left[\int^{x = \infty}_{x_i} \rho(r) \epsilon(\lambda) [\displaystyle \rm exp \mit \{hc/\lambda kT(r)\} - 1]^{-1} dx\right] }
\end{equation}
\begin{equation}
\rm where \; \mit x_i = \;\; ^{\displaystyle \rm 0 \;\;\;\;\;\:\;\;\;\;\;\;\;\;\;  if \;\;\; \mit  r > r_i}_{\displaystyle  \sqrt{r_i^2 - b^2} \;\;\; \rm if \;\;\; \mit r < r_i}
\end{equation}
 
\noindent and $x$ is the distance along the line of sight (see Figure
3).  The integration begins at $x_i = 0$ for $r > r_i$ and $x_i =
\sqrt{r_i^2 - b^2}$ for $r \leq r_i$ (i.e. the value of $x$ at the
intersection of the line of sight with the inner radius of the shell).
The shell starts at an inner radius $r=r_i$, but the integration is
terminated at $x = 8\arcsec$, as continuing the integration beyond
this point results in no additional flux to 1 part in $10^6$ in the models considered.  In (1),
$p$ is the fraction of the total flux due to extended emission and
$I_{\lambda}(b)_{phot}$ is the photospheric flux distribution, defined

\begin{equation}
I_{\lambda}(b)_{\scriptstyle phot \mit} = ^{\displaystyle(1 - p) I_{\scriptstyle \lambda _{\scriptscriptstyle  \rm TOT \mit}} \displaystyle \rm \;\;\; if \;\;\; \mit b = \rm 0 \mit}_{\displaystyle \; \rm 0 \;\;\;\;\;\;\;\;\;\;\;\;\;\;\;\;\;\;\; if \;\;\; 
\mit b > \rm 0 \mit}
\end{equation}

Performing this integration numerically for an assumed form of $\rho (r)$, we generated a profile of the
dust shell emission as a function of the mean radius $b$.  The model
was then normalized to the shell's total flux ($pI_{\lambda _{TOT}}$)
and added to the point source ($I_{\lambda}(b)_{\rm phot}$).  To
include observational effects, we converted this one-dimensional
profile into a two-dimensional image, then convolved the image with
the $\alpha$~Tau point spread function.  
Lastly, we extracted azimuthally averaged one-dimensional profiles 
$I_{\lambda}(b)_{\rm CM \mit}$ (CM denotes the convolved model) from
the model image for comparison with similar profiles extracted from 
from the original mosaics.

The models were fitted to the data using a downhill simplex
minimization method (\cite{NR}).  The simplex method attempts to find
the best model by minimizing $\chi ^2 = \displaystyle \sum _b
\textstyle \frac{1}{\sigma ^2} (1 - \frac{I_{\lambda}(b)_{\rm CM
\mit}}{I_{\lambda}(b)_{\rm observed}})$.  Initially, the
uncertainty in each point was given by $\sigma = \rm constant$, but
subsequently we tested several other possible definitions of the
uncertainty $\sigma$ by allowing $\sigma$ to vary as a function of
radius ($\sigma = \sigma (b)$).  For each of these cases, however, we
found that the parameters returned from the fits were similar for each
of our different definitions of $\sigma$.  We therefore conclude that
the fitted parameters are robust against both the model details and
the boundary conditions.

\subsection{Power-Law Models}

Initially, the data were fitted using a simple power-law for the dust 
density ($\rho \propto r^{-n}$), with an inner radius cutoff.  Two models were fitted 
to the 11.7 and 17.9~\micron\ data separately, and a third model linked 
the two wavelengths by using common values for the temperature, 
exponent, and inner radius, but separate flux percentages, $p_{11.7}$ and 
$p_{17.9}$.  The best-fit values of parameters $r_i$, $T_i$ 
(the temperature at the inner radius of the dust shell), $n$, and 
$p$ ($p_{11.7}$ and $p_{17.9}$ for the two-color model) for these 
models are listed in Table 1.  

The observed and modeled 1--D profiles of the star + shell are plotted
in Figure~4.  The two-color power law model fits the data quite well,
with a $\chi ^2$ per degree of freedom value of 1.2 assuming a
constant 5\% error on all data points.  The power-law model has a
serious weakness, however: it predicts a value of 0$\farcs$07 for the
inner radius of the shell, in contradiction to values of $\sim
1\arcsec$ indicated by interferometric observations (Danchi et
al. 1994).

\subsection{The Spike Models}

To resolve the incompatability of our power law models with previous
observations, we tested two thin-shell models.  The first model, the
spike model, assumed a thin shell with thickness of only $0\farcs 1$.
This model did not match the observational data nearly as well as the
initial power-law model.  The $\chi ^2$ value was much larger than
that found from the thick shell model (by a factor of roughly 100),
and the resulting parameter values (see Table 1), such as $r_i =
4\farcs 5$, are not reasonable, as well as incompatible with the model
suggested by Danchi et al (1994).  Therefore, we rule out the basic
spike model.

The second thin-shell model (hereafter termed the spike+ model)
was a combination of the spike and extended shell power-law models.  
We assumed that both the thin and extended shells began at the same inner 
radius, $r_i$, and that the temperatures of the two components were equal 
at this radius. One additional parameter $f_p$ was introduced to
specify the ratio of the density of the thin shell to the density of 
the extended material 
($f_{\rho} = \rho _{\rm thin}/\rho_{\rm thick}$), 
giving us $\rho (r) = \rho _{\rm thick}(f_p \delta (r_i) + r^{-n})$.  
This model, like the power-law model, was fitted to the 11.7\micron \
and 17.9\micron\ data separately, then was run with the two wavelengths
linked by common values for $T_o$, $r_i$, $n$, and $f_p$.  Numerical
results from these models are listed in Table~1 and the azimuthally
averaged 1--D profiles are included in Figure~4.  The spike+ model also
produces a good fit to the observed data, with a $\chi ^2$ value of
1.0 per degree of freedom for a uniform 5\% error on data points.  In addition, the model
returned inner radii $r_i \approx 1\arcsec$, in good agreement     
with the interferometric observations.

\subsection{2--D Models}

One of the major assumptions made in the model calculations was that
the circumstellar dust exists in a spherical shell which starts at a
well-defined inner radius and extends to a significant distance from
the star.  In order to test this assumption, we also tried a model
with a disk geometry.  Using IRAF software, we measured the
eccentricities of both $\alpha$~Ori and $\alpha$~Tau to be $\leq 0.05$
at 11.7~\micron \ and $\leq 0.1$ at 17.9 \micron, so a disk would have
to be nearly face-on.  We again fit the disk model at each observed
wavelength, and then on both wavelengths simultaneously.  Both the
single- and two-color models were marginally worse than our spherical
shell models (Table 2), with a $\chi ^2$ value of 1.4 per degree of
freedom.  This model is included solely for completeness; since a
number of evolved stars display spherically symmetric dust shells, and
disk distributions are not observed in these objects, we believe that
the disk model is not a realistic representation.

\section{Results and Conclusions}

We found that the two-color power-law and and the two-color spike+
models fit the $\alpha$~Ori images about equally well (Figure 4).
However, because of the disagreement between the power-law models and
previous interferometric observations, we feel that the spike+ model
better represents the $\alpha$~Ori shell. To examine the shape and
character of the modeled dust distributions in more detail, we created
PSF-subtracted images using the flux ratios determined by the models.
Plots of the mean surface brightness of the dust shell component as a
function of radius for both models (the two differ in scaling of the
subtracted image of $\alpha$~Tau) at both 11.7 and 17.9~\micron\ are
shown in Figure 5.

The ratio of the 11.7\micron \ flux to the 17.9\micron \ flux,
multiplied by the ratio of the emissivities of the silicate dust
grains at these two wavelengths (Draine and Lee 1984), indicates the
grain color temperature as a function of radius.  The temperature
results for both the power-law and spike+ models are presented in
Figure 6.  Also shown are the effective color temperature profiles of
the dust distributions of our best-fit models, calculated numerically
using the $\rho (r)$ density distributions and the temperature profile for dust grains in thermal equilibrium, $T = T_o(r_o/r)^{1/2}$.

From our two-color spike+ model fits, we find that the dust shell
provides 33\% $\pm$ 3\% of the total flux at 11.7 $\mu$m and 45\%
$\pm$ 4\% at 17.9 $\mu$m.  From the value $f_p$, the ratio of the
density of the thin shell at $r_i$ to the density of the extended
material at $r_i$ is $\rho _{\rm thin} \mit \geq (75 \pm 25)\rho
_{thick}$ (we know that $\rho _{\rm thin}\Delta r \simeq$ constant,
where $\Delta r$ is the thickness of the thin shell, and are therefore
limited by the size of our resolution element).  Because of the
difference in the extent of the two portions of the model dust
distribution, the thin shell component provides only $\sim$40\% of the
dust shell flux (13\% of the total flux) at 11.7\micron \ and only
$\sim$30\% of the flux (14\% of the total flux) at 17.9\micron.

We also find a value for the inner shell radius of $r_i = 1\farcs 0
\pm 0\farcs 1$.  This, unlike our simpler power-law model, agrees well
with previous interferometric observations (Bester et al. 1991).
While this agreement leads us to believe that this model is a better
depiction of the actual dust distribution, the dust grain color
temperature at $r_i$ is only 460 K, well below the condensation
temperature of $\sim$ 1500 K for silicates (Salpeter 1974).  One
possible explanation for the low temperature is that a stellar wind
has pushed the dust inside one arcsecond into the thin shell now
observed.  The dust presumably condensed at a much smaller radius,
then cooled as it was driven away from the star.  This would
presuppose that the star has not recently been undergoing heavy
mass-loss, so that no new dust has condensed near the star.  If this
scenario is accurate, then we can estimate how long it has been since
the last major mass-loss event.  The last major variation in the
brightness of $\alpha$~Ori occurred during the period 1941-1945.  If
this brightness variation corresponds to the last major mass-loss
event, the mean velocity required to bring the dust from this event to
1$\arcsec$ is 12 km s$^{-1}$. This velocity is in agreement with the
11 km s$^{-1}$ CO linewidths.  It should be mentioned that previous
work (Danchi et al. 1994) modeled the circumstellar material as a
single thin shell (or possible two thin shells), wheras our work
indicates that a significant portion of the dust emission comes from a
very large extended shell.  This is not a surprising difference, given
that the interferometric observations were taken with long baselines,
making the observations very insensitive to flux coming from large
radii.

While we do not favor the power-law model, it is interesting to note
that the power-law model color temperature at the inner shell radius
($r_i = 0\farcs 1$) corresponds well to the condensation temperature
for silicate grains.  By balancing the flux absorbed by an individual
grain with the flux emitted by the grain, we find that reasonable
grain radii ($a = 0.01\micron$ and $a = 0.1\micron$) produce
temperatures ($T_{dust}(r_i) = 1740$ K and $T_{dust}(r_i) = 1890$ K
respectively) which agree with both silicate condensation and our
power-law model.

The temperature of the dust at one arcsec from the star (the inner
shell radius) is found to be $T_o = 460 \pm 20$K.  Examining the
spike+ model plots on Figure 6, we see that this temperature is in
good agreement with the predicted color temperatures from the $\rho
(r)$ distribution, especially for projected radii larger than 1$\arcsec$.  We
find an average temperature for the spike+ model of 205 $\pm$ 35 K,
compared to $T_{\rm mean} \approx 300$ of Bester et al. (1991).

The last parameter of the dust shell model which was returned from our
minimization method was the exponent of the density dependence upon
radius for the extended emission, $n$.  The value of this exponent
from our best-fit spike+ model is $n = 0.9 \pm 0.1$.  However, $n = 2$
for a shell expanding at constant velocity, and $n = 3/2$ for a shell
expanding at the escape velocity.  Therefore, any plausible,
spherically symmetric, steady state density distribution should have
$3/2 \leq n \leq 2$.  From this discrepancy, we conclude that the
circumstellar dust flow must be episodic.  This is not surprising,
given that the inner radius of the shell also suggests episodic
emission.

\acknowledgements

The authors would like to thank the staff at Palomar Observatory for
their assistance and the referee for helpful comments.  Figures 1 and
2 can be found in gif format at
http://www.people.cornell.edu/pages/sr22/Aori. This research was
partly supported by NASA Contract 960803.

\newpage

\begin{figure}
\caption{Images at 11.7$\micron$\ of $\alpha$ Ori (a), $\alpha$ Tau (b), and the subtracted image (d) produced with the PSF scaled to 39\% of the total flux in the $\alpha$~Ori image.  Also presented is a plot of the total flux observed within annuli of inner radius $b$ (c).  Each of the images are 13$\farcs$3 on a side; the small bar in the upper right of each image is 1$\arcsec$ long.}
\end{figure}

\begin{figure}
\caption{Images at 17.9$\micron$\ of $\alpha$ Ori (a), $\alpha$ Tau (b), and the subtracted image (d) produced with the PSF scaled to 49\% of the total flux in the $\alpha$~Ori image.  Also presented is a plot of the total flux observed within annuli of inner radius $b$ (c).  Each of the images are 13$\farcs$3 on a side; the small bar in the upper right of each image is 1$\arcsec$ long.}
\end{figure}

\begin{figure}
\caption{A diagram of the physical geometry used for our models.}
\end{figure}

\begin{figure}
\caption{Plots of the power emitted by annular rings of $\alpha$ Ori (including the star and the shell), with the best-fit model overlaid (a) for the power-law model at 11.7 $\mu$m, (b) for the power-law model at 17.9 $\mu$m, (c) for the spike+ model at 11
.7 \micron \ and (d) for the spike+ model at 17.9 \micron \ .  Units used in these plots are arbitrary.}
\end{figure}

\begin{figure}
\caption{Plots of the mean flux from the dust shell, produced by subtracting a scaled PSF from the raw image of $\alpha$~Ori, as a function of distance from the star (a) for 11.7 $\mu$m and (b) for 17.9 $\mu$m.  Also provided is a plot of 25\% of the flux
 of $\alpha$ Tau.  Examination of these plots shows that at small radii ($b < 1\arcsec$), the implied shell brightness is highly model-dependent, while at large radii ($b > 1\arcsec$) the shell brightness is very similar for both models.}
\end{figure}

\begin{figure}
\caption{The color temperature of the dust shell, using a two-point blackbody fit from the PSF-subtracted images.  Also plotted are the predicted effective color temperature curves for each model.}
\end{figure}

\clearpage

\begin{table}
\begin{center}
\begin{footnotesize}
\begin{tabular}{llllll}
Parameter  & Model    & Spike+ & Power-law\tablenotemark{a} &  Disk\tablenotemark{a} & Spike\tablenotemark{a} \\\tableline
T$_i$\tablenotemark{b} &&&&& \\
& 11.7 & 445$\pm$40 K & 930$\pm$80 K  & 750 K & \nodata \\
& 17.9 & 510 (+70/-50) K & 1800 (+1300/-700) K & 790 K & \nodata \\
& Two-Color & 460$\pm$20 K  & 1700$\pm$700 K  & 580 K & 490 K \\
r$_i$\tablenotemark{c} &&&&& \\
& 11.7 & $1\farcs 1\pm 0\farcs 1$ & $0\farcs 66\pm 0\farcs 1$ & $0\farcs 58$ & \nodata \\
& 17.9 & $0\farcs 9 \pm 0\farcs 1$ & $0\farcs 15 (+0\farcs 2/-0\farcs 1)$  & $0\farcs 42$ & \nodata \\
& Two-Color & $1\farcs 0 \pm 0\farcs 1$ & $0\farcs 07 (+0\farcs 3/-0\farcs 06)$ & $0\farcs 49$ & $4\farcs 9$ \\
n\tablenotemark{d} &&&&& \\
& 11.7 & 0.93 (+0.09/-0.06) & 2.0 (+0.2/-0.1) & 0.8 & \nodata \\
& 17.9 & 0.93$\pm$0.13 & 1.4$\pm$0.1& 0.22 & \nodata \\
& Two-Color & 0.9$\pm$0.1 & 1.0$\pm$0.1 & 0.01 & 2.1 \\
p$_{11.7}$\tablenotemark{e} &&&&& \\
& 11.7 & 0.33 (+0.03/-0.02) & 0.35 (+0.03/-0.02) & 0.33 & \nodata \\
& Two-Color & 0.33$\pm$0.03 & 0.39 (+0.03/-0.05) & 0.30 & 0.05 \\
p$_{17.9}$\tablenotemark{f} &&&&& \\
& 17.9 & 0.41$\pm$0.04 & 0.47 (+0.03/-0.04) & 0.40 & \nodata \\
& Two-Color & 0.45$\pm$0.04 & 0.49$\pm$0.05  & 0.41  & 0.16 \\
f$_p$\tablenotemark{g} &&&&& \\
& 11.7 & 0.986$\pm$0.003 & \nodata & \nodata & \nodata \\
& 17.9 & 0.976 (+0.01/-0.06) & \nodata & \nodata & \nodata \\
& Two-Color & 0.986$\pm$0.04 & \nodata & \nodata & \nodata \\
$\chi ^2$&&&&& \\
& Two-Color & 1.0 & 1.2 & 1.4 & \nodata \\
\end{tabular}
\end{footnotesize}
\end{center}
\tablenotetext{a}{These models are included solely for completeness.}
\tablenotetext{b}{The temperature of the dust at the inner edge of the shell.}
\tablenotetext{c}{The inner radius of the shell.}
\tablenotetext{d}{The density power-law exponent.}
\tablenotetext{e}{The fraction of the total $\alpha$~Ori flux from the dust shell at 11.7$\micron$.}
\tablenotetext{f}{The fraction of the total $\alpha$~Ori flux from the dust shell at 17.9$\micron$.}
\tablenotetext{g}{The ratio of the density of the thin shell to the density of the extended shell.}
\end{table}

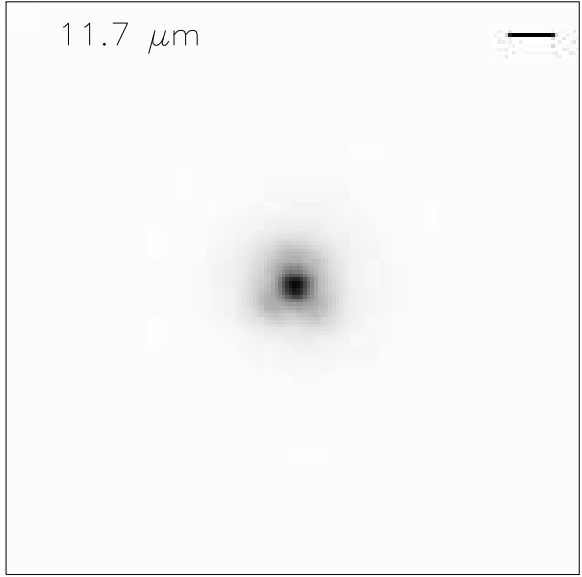
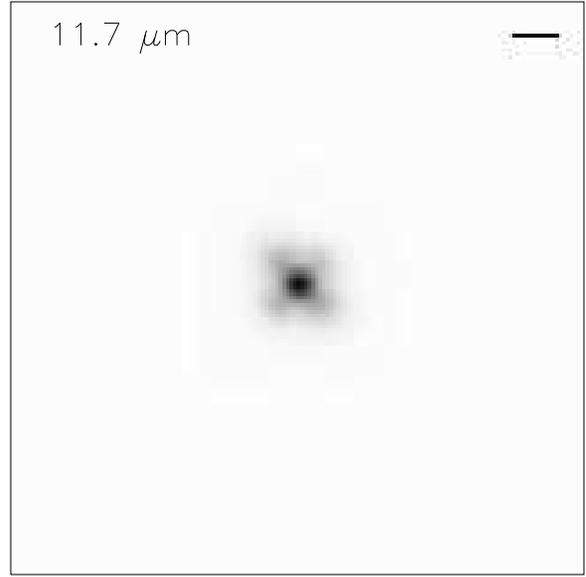
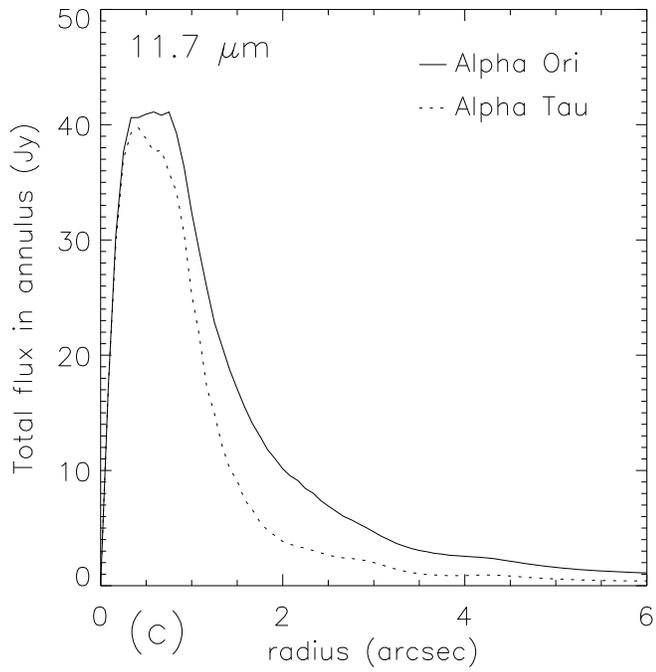
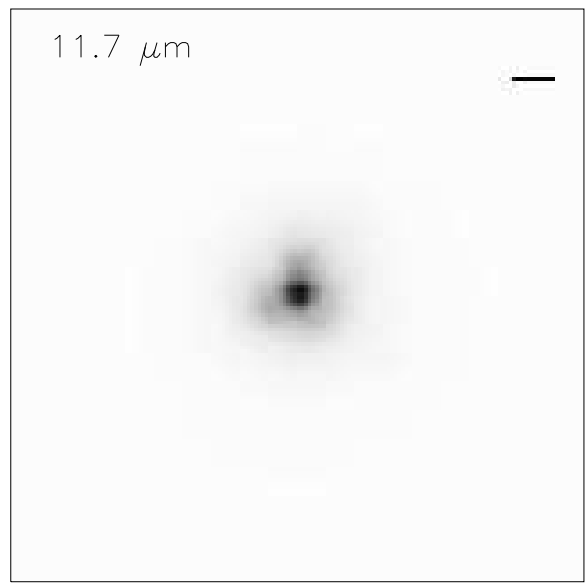
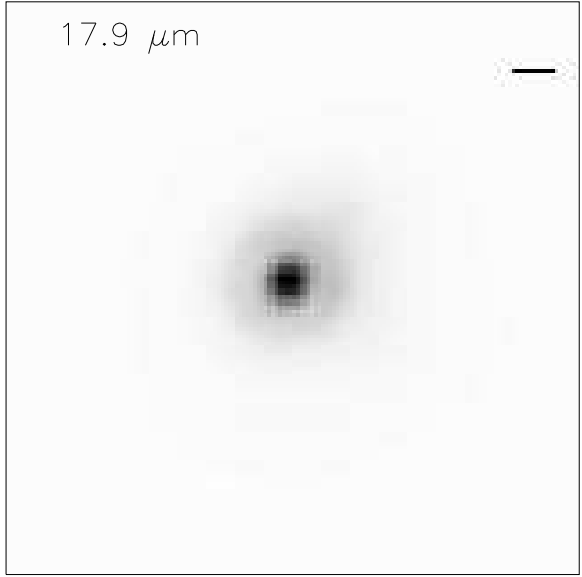
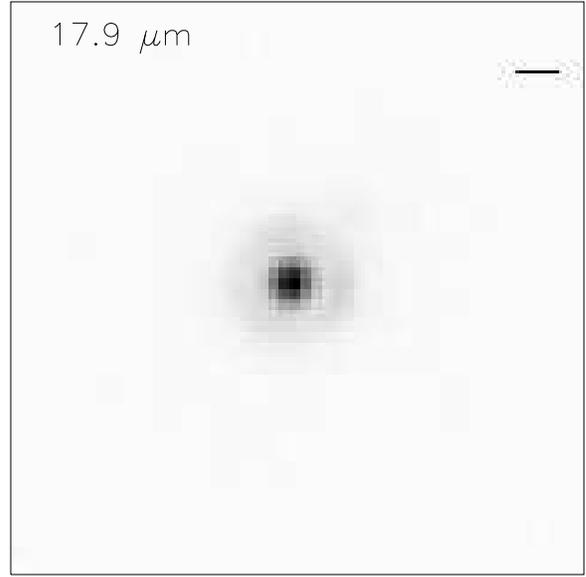
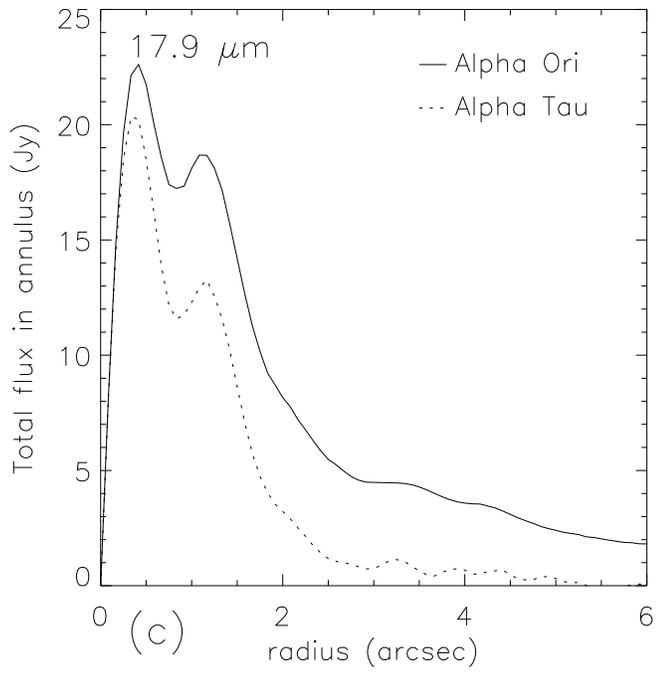
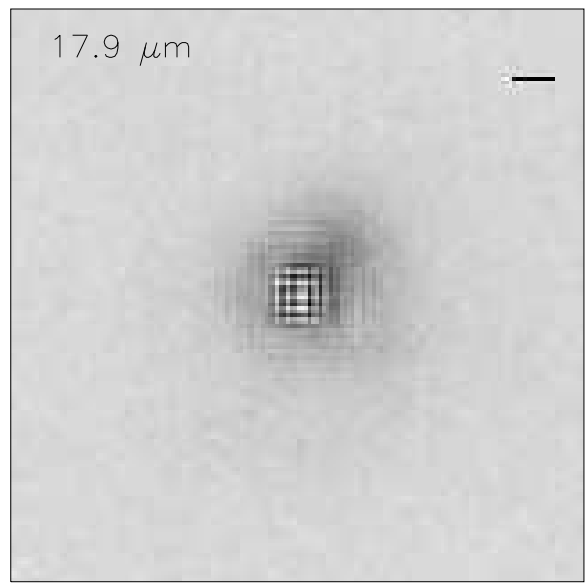
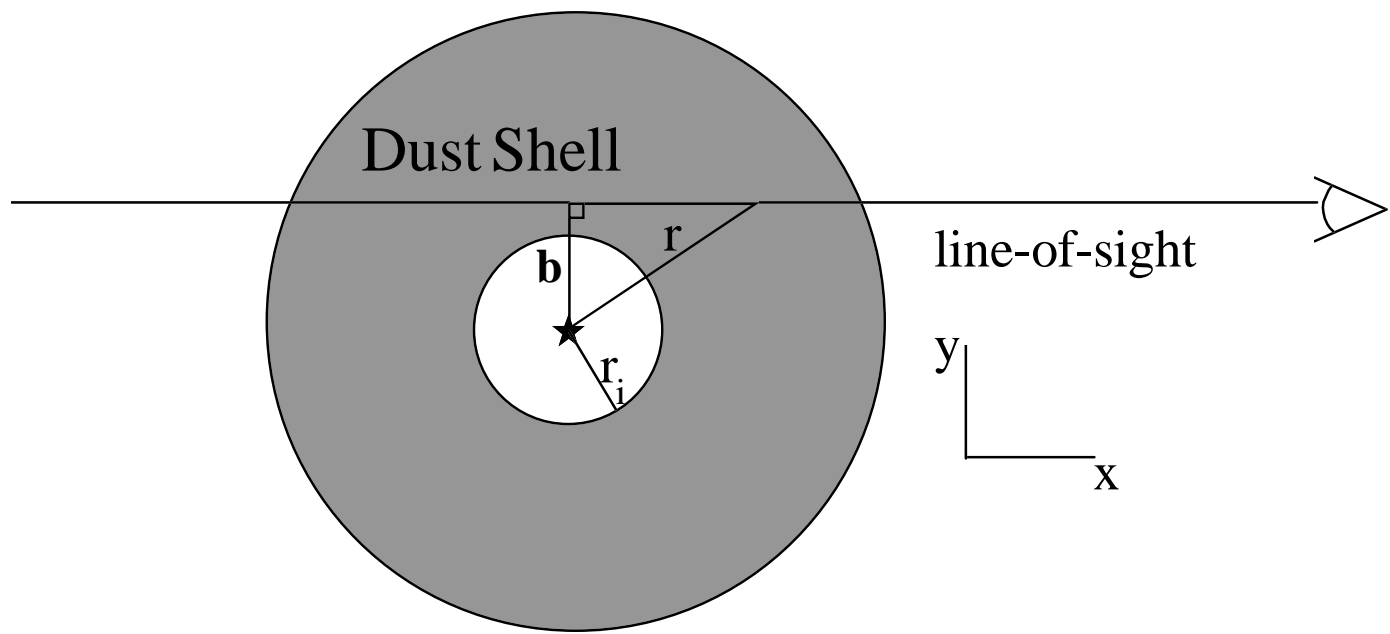
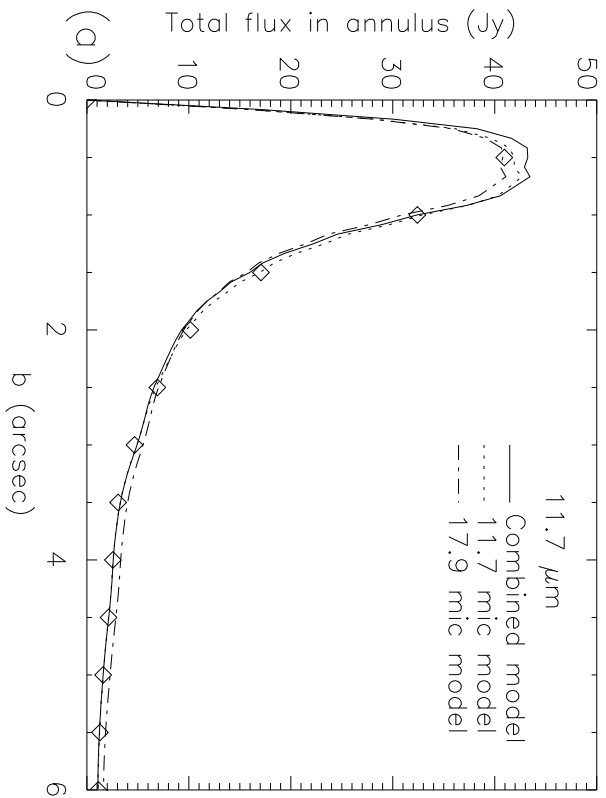
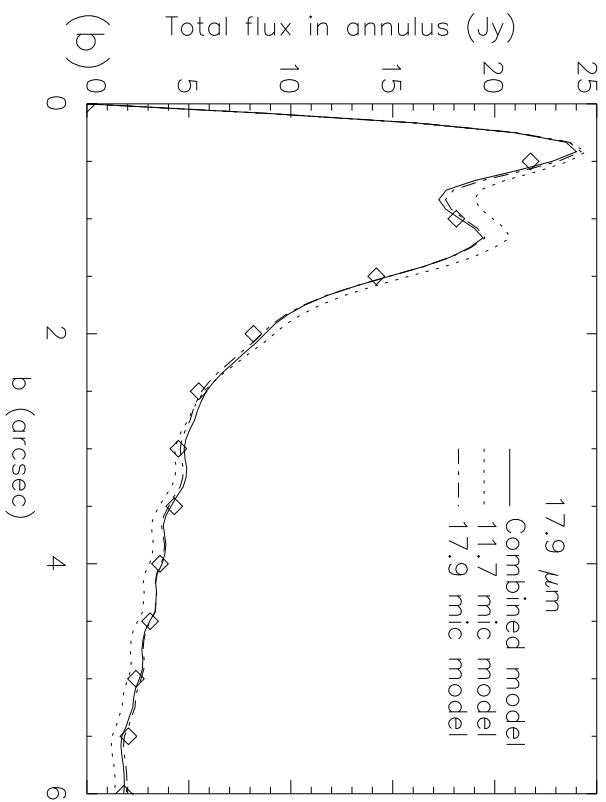
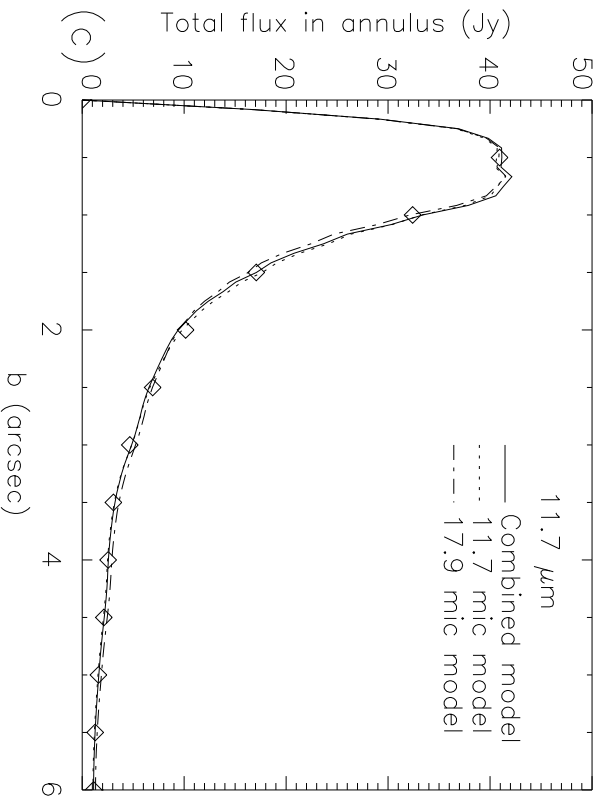
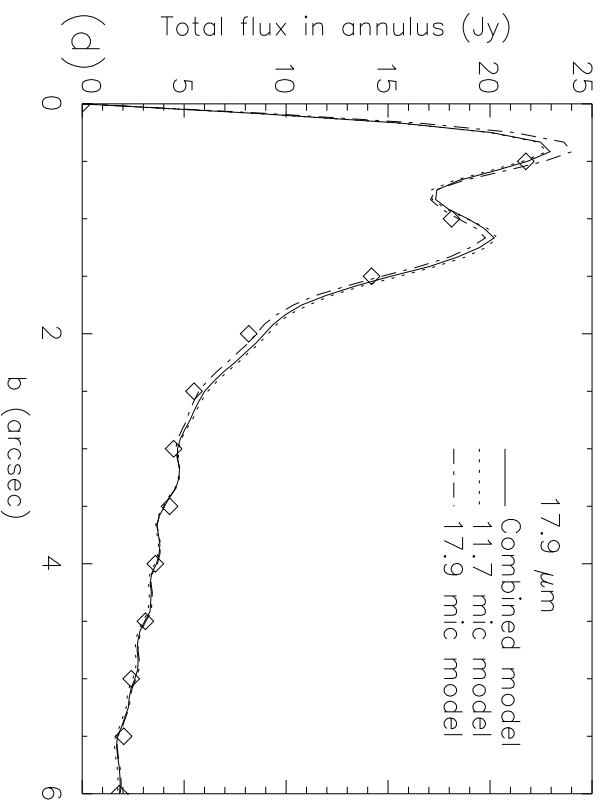
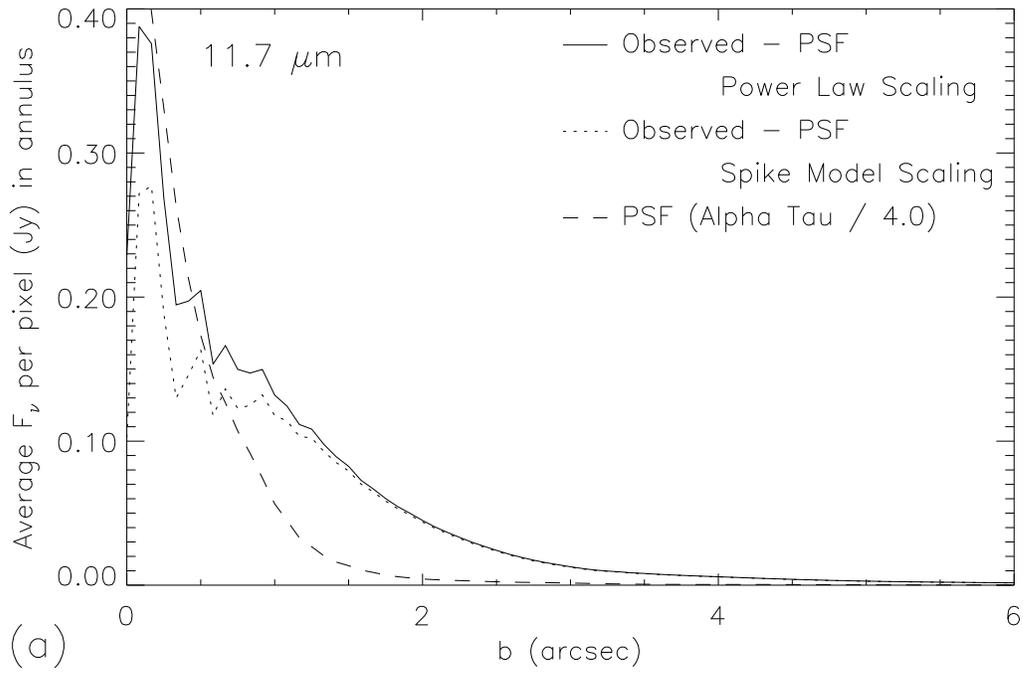
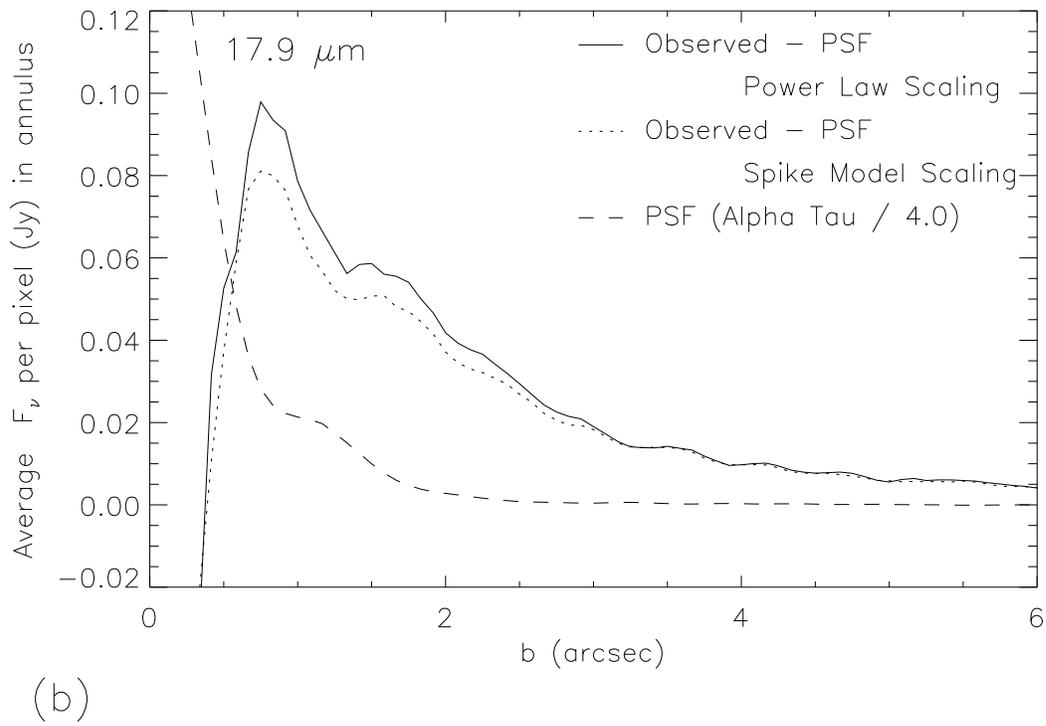
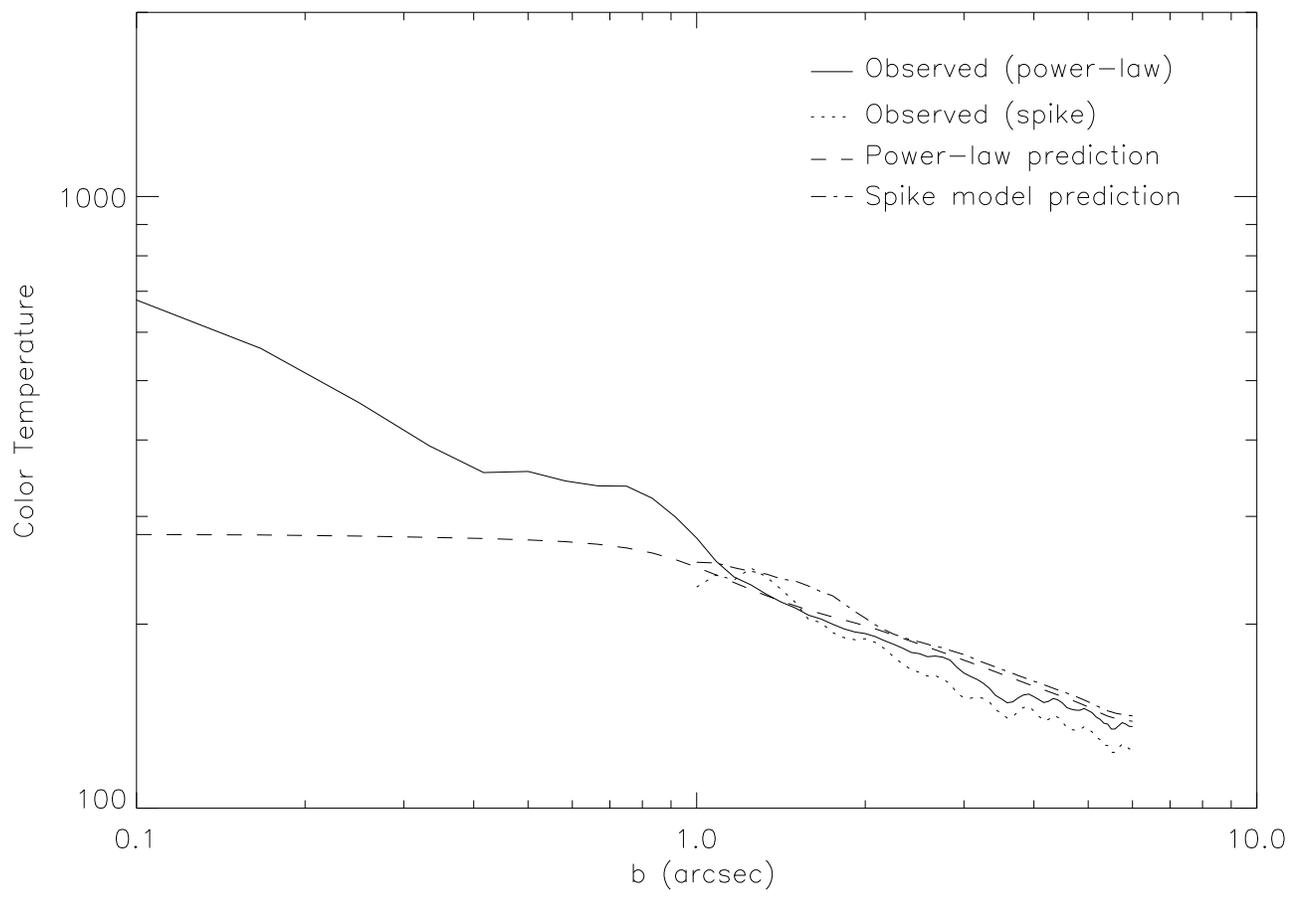
\end{document}